# Robust multi-scale multi-feature deep learning for atomic and defect identification in Scanning Tunneling Microscopy on H-Si(100) 2x1 surface

Maxim Ziatdinov,[1] Udi Fuchs,[2] James H.G. Owen[2], John N. Randall,[2,a] and Sergei V. Kalinin[1,b]

[1] The Center for Nanophase Materials Sciences, Oak Ridge National Laboratory, Oak Ridge, TN 37831

[2] Zyvex Labs LLC, 1301 North Plano Road, Richardson, Texas 75081

The nature of the atomic defects on the hydrogen passivated Si (100) surface is analyzed using deep learning and scanning tunneling microscopy (STM). A robust deep learning framework capable of identifying atomic species, defects, in the presence of non-resolved contaminates, step edges, and noise is developed. The automated workflow, based on the combination of several networks for image assessment, atom-finding and defect finding, is developed to perform the analysis at different levels of description and is deployed on an operational STM platform. This is further extended to unsupervised classification of the extracted defects using the mean-shift clustering algorithm, which utilizes features automatically engineered from the combined output of neural networks. This combined approach allows identification of localized and extended defects on the topographically non-uniform surfaces or real materials. Our approach is universal in nature and can be applied to other surfaces for building comprehensive libraries of atomic defects in quantum materials.

---

[a] jrandall@zyvexlabs.com
[b] sergei2@ornl.gov

Atom by atom fabrication is one of the longest-held targets of nanoscience and nanotechnology. Originally envisioned by luminaries such as Feynman[1] and Drexler[2] as a purely theoretical concept, it was brought into the realm of the possible in the seminal work by Don Eigler[3]. The demonstration of single atom positioning and writing has firmly brought nanoscience to the attention to the scientific and broad community and contributed to the launch of the National Nanotechnology Initiative in the US[4] and equivalent programs in multiple countries around the globe.

The initial proof of concept of atomic manipulation has led to a series of fundamental studies of new electronic functionalities enabled by this atomic based construction, including quantum corrals[5] and holograms[6], molecular cascade arrays[7], atomic scale information storage[8-10], magnetoelectronic and spintronic structures[11,12], and many others. The celebrated "The Boy and his atom" movie illustrates the outstanding level of control over scanning tunneling microscopy (STM) based atomic manipulation[13]. However, since the time of the original Eigler experiments and until very recently, the attention of the scientific community has been focused on fundamental research, with the enabling instrumentation being available only in a small number of facilities worldwide.

This situation has changed completely over the last several years. The progress of the semiconductor roadmap at this point leads to the commercialization of sub-10-nm technologies, and exploration of single-digit Si devices, with the associated needs for fabrication and metrology. Perhaps even more importantly, quantum computing and quantum information systems are moving to the front of research and development, necessitating the development of corresponding fabrication and metrological tools. The signing of the National Quantum Initiative in December 2018 by the United States government is dramatically boosting funding for academic and national laboratory quantum research. The subsequent formation of the Quantum Economic Development Consortium has demonstrated significant industrial interest in research for quantum technologies. While most of the industrial effort has until now been focused on the Josephson junction[14] and trapped ion/atom qubit systems[15], promising developments and proof-of-concept results have been obtained with solid-state qubits such as P/Si, as has been demonstrated by groups in the University of New South Wales,[16,17] Sandia National Lab [8], and National Institute for Standards and Technology [19].

However, the key roadblock towards the broad implementation of solid-state qubits and other single- and several atom devices via STM manufacturing is the issue of automatic control and feedback-based operation. Until recently, STM based fabrication relied on the known functional response of a material to the sequence of operational steps, much like the resist-based process in classical semiconductor nanofabrication The current process provides no feedback to the system other than human observation and action. Correspondingly, instrumental drift, spurious reactivity (e.g. formation of an adatom group, etc.) all precluded successful fabrication unless there is a diligent and expert human operator. Notably, similar problems have recently emerged in the context of Scanning Transmission Electron Microscopy (STEM) based atomic manipulation.[20-22]

An alternative approach is offered by the incorporation of image-based feedback into the real-time microscope operation. Here, the specific set of manipulation and control steps is being made based on the information on the position of atomic species determined within a scan (or its parts) determined in real time. This, in turn, necessitates the development of image analysis tools capable of identification and semantic segmentation of images. Previously, this approach was demonstrated for STEM-based crystallization, utilizing the intensity of the Fourier transform peak as the feedback signal.[23] However, the applications in atomic based STM and STEM fabrication both require the identification of individual species, rather than merely periodicities, in the presence of noise and instrumental drift.

Previously, several groups have demonstrated the application of deep neural networks for fast and automated identification of the type and position of atoms and atomic columns as well as point-like structural irregularities (atomic defects) in static and dynamic STEM data[24-28]. Specifically, a combination of deep neural networks with domain-specific knowledge allowed to reconstruct the reaction pathways for point defects in 2D materials[26], trace the structural evolution of atomic species during the electron beam manipulation[24,26], and create a library of defect configurations in 2D materials such as graphene.[27,28] More recently, the application of deep learning for STM tip conditioning[29], as well as a supervised classification of defects and subsequent identification of clean surface regions for atomic fabrication[30] have been demonstrated. However, incorporation of the deep learning networks both for the implementation of automated STM (or STEM) experiments and fundamental physics studies necessitates robustness towards non-atomically resolved regions, extended defects, step edges, etc. Note that while in post-acquisition data analysis such robust approaches generally accelerate analysis workflow and obviate the need for human intervention (already significant for non-trivial data volumes possible with modern platforms), for automated experimentation and especially STM based lithography[31], high robustness is an absolute prerequisite since any misidentification can result in incorrect operation with potential deleterious consequences for the patterned structure or even the microscope.

Here we have developed a robust approach for locating atoms and defects from the STM images in the presence of extended and non-atomically resolved images. We define our approach as weakly-supervised machine learning as the only information about materials structure that we provide to the algorithm is i) that there are step edges and terraces, ii) that the surface represents rows of Si dimers, iii) that there are defects on a surface where the Si dimer rows are interrupted. However, we do not provide any specific prior knowledge about the defects themselves, allowing the algorithm to discover their classes in an unsupervised (and unbiased) manner. This approach is implemented as a workflow of three functionally different deep learning networks performing different stages of analysis. We demonstrate this approach for the unsupervised classification of surface defects on the Si(100)-surface, providing insight into the chemistry of the surface.

As a model system, we have chosen the H-terminated Si(001) surface, which is prepared by a standard process of flashing to 1250°C to create a clean, atomically-flat surface, followed by H termination with atomic H generated by a hot filament. The initial surface, as shown in Fig.1 a and b, comprises primarily rows of silicon dimers, terminated with H atoms. Both bright and dark defects are seen on this surface. Dark defects (vacancies) result from missing Si atoms, and are physical depressions in the surface. Bright defects are either missing H atoms, which show up bright due to their electronic structure, or are other adsorbates, giving a physical bump on the surface (adatoms). STM Lithography can be used to remove H atoms from this surface. The small bright patches in Fig. 1c and d, are examples where a few dimers have been stripped on H. To do this, the tip must be carefully aligned to the dimer row, the tip conditions are set to +4.5V, 4 nA, and then moved slowly over the area to be depassivated. For dopant atom placement, these patches need to be exactly 3 dimers, and to be an exact distance from other features on the surface. Currently, this patterning is performed manually, but by identification of the dimer row locations, and the position of position references, this works hopes to automate this process, and improve the speed and precision of the pattern writing.

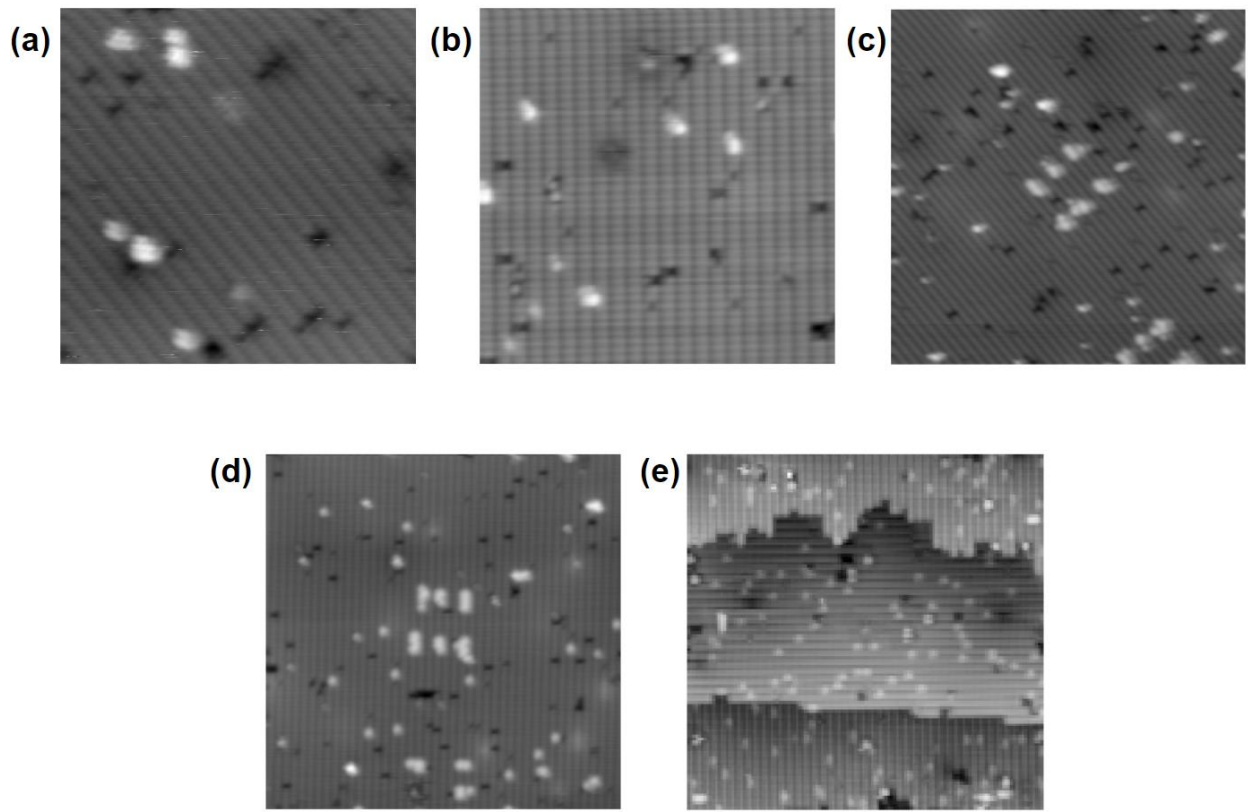

**FIGURE 1.** Overview of experimental STM data from Si(100) surface. The imaging conditions are: $V_{bias}$ = -2.5 V, $I_{setpoint}$ = 0.2 nA. The scan size is 24 nm$^2$ in (a, b) and 32 nm$^2$ in (c-e). See the text for more details.

Several representative examples of the surface structures observed on the Si(100) surfaces are shown in Figure 1. Several aspects of these images make them extremely complex for image recognition. The images contain a large number of point defects of the adatom and vacancy types, many of which are located next to each other and form clusters. This tendency towards cluster formation precludes the use of a simple image analytic tool for defect separation and identification. Secondly, while the dimer rows are generally well visible (and conversely if rows are not resolved the tip state should be optimized), the atomic resolution within the rows is generally difficult to detect. Finally, the images may contain a number of step edges or non-atomically resolved features.

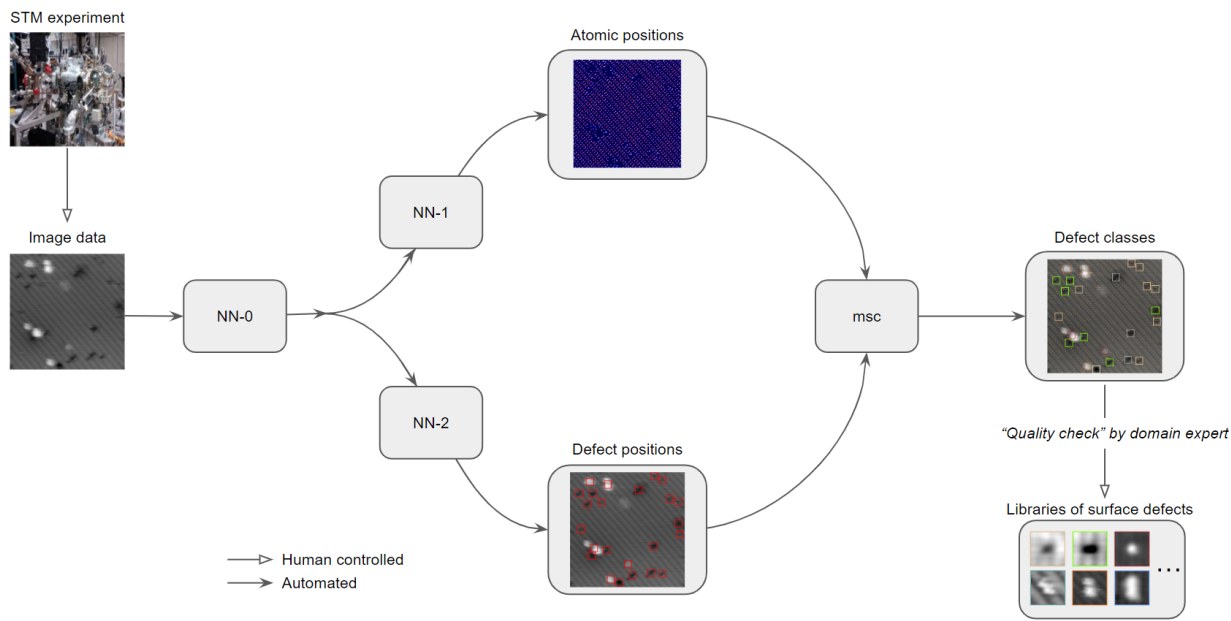

**Figure 2.** Schematic workflow for the analysis of experimental STM data. Images are passed through fully convolutional neural networks tasked with separating different terraces (or, if an image contains only a single terrace, passing it to the next networks) (NN-0), finding all the atoms (NN-1) and with locating the surface defects (NN-2). The outputs of the last two networks are then merged together and used to construct features associated with geometric shape and apparent height of the extracted defects. These features are passed to a mean-shift clustering algorithm (msc) that categorizes defects into different classes in an unsupervised fashion. In the end, the labels are refined by a domain expert and used to create a library of surface structures and defects that can be reused for other classification tasks and automated experiments (in this case, the 'msc' part in this scheme will be replaced by a supervised classifier).

Here, we propose and develop a deep learning workflow that enables robust multiscale analysis of such data sets. The first step towards the implementation of automatic image-based feedback and fabrication is the development of robust algorithms capable of fast identification of individual atomic objects. This is achieved by creating a workflow that combines multiple deep learning networks with different computer vision tools in the pre- and post-processing steps. All the neural networks are the fully convolutional deep learning networks with an encoder-decoder type of architecture,[32] which earlier proved to be successful in finding atoms and defects in electron microscopy images.[25-28] The major differences are the type of training data and the post-processing algorithms applied to an output of a neural network. We found that, due to a classical in machine learning tradeoff between accuracy and generalization, having multiple "specialized" networks leads to a higher reliability compared to having a single "general" neural network.

We start by separating data with step edges and terraces via a fully convolutional neural network model (NN-0) followed by the connected components labeling algorithm. The model is trained using experimental data where step edges and terraces were labeled on the pixel level using a combination of a Canny edge detector for basic edge detection, Gaussian processes to recover missing parts of the detected edges and blob filtering to remove small blobs associated with edges of small surface defects, and further augmented by adding noise (Gaussian, Poisson and salt-and-pepper), zooming-in, resizing, and horizontal/vertical flipping to account for different acquisition parameters, different orientation of edges, etc. As an output, the network fragments an STM image into individual sub-images of terraces and steps and use those sub-images for the subsequent atom finding and defect classification (Figure 3).

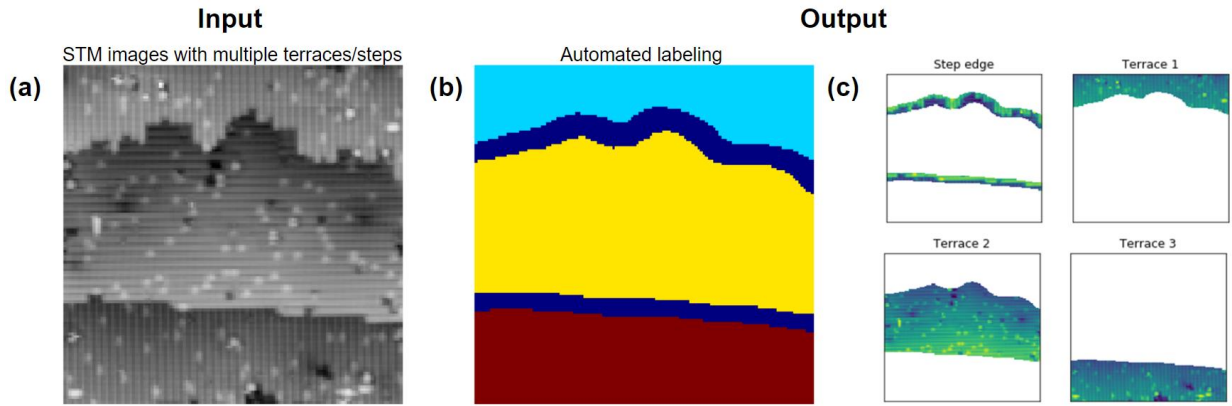

**Figure 3.** Automated detection and labeling of step edges and terraces from the experimental STM image. (a) Experimental STM image. (b, c) The output of our model, which automatically labels steps and terraces (b) in the experimental image and returns multiple images of individual terraces and step edges (c).

The images with terraces are then passed through two different fully convolutional neural networks: network-1 (NN-1), which is tasked with finding Si surface atoms and network-2 (NN-2), which is tasked with locating defects on the surface (Fig. 2 and Fig. 4). The purpose of NN-1 is to automatically identify the dimer atoms on the terraces while avoiding surface regions where the atomic structure is ambiguous, whereas the purpose of NN-2 is the automated identifications of point/atomic defects. The NN-1 was trained on the simulated data of Si rows with vacancies and "protrusions". The NN-2 was trained using a few labeled experimental images to identify defects that break the periodicity of the surface dimerized structure. To deal with the noise in the experimental data, as well as to better account for the fact that the Si lattice is periodic, we replaced regular convolutions in the middle of both networks with a spatial pyramid of dilated convolutions with dilation rates {2, 4, 6}. Both networks can work with input images of variable size as long as the image width and height are divisible by $2^n$, where $n$ is a number of max-pooling operations, which for the NN-1 and NN-2 model is equal to 3 and 1, respectively. There is, however, an optimal pixel-to-angstrom ratio for which one can get the most robust and accurate results from the network. It currently takes ~ 0.01 s on a standard GPU to obtain atomic or defects positions from the raw experimental image of 256 x 256 resolution. The atom finder neural network has been successfully deployed on to an STM system at Zyvex Labs.

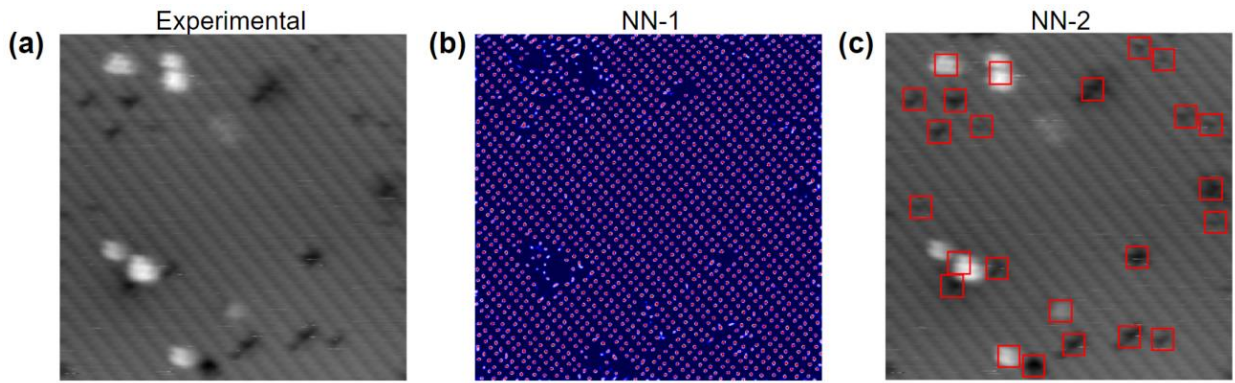

**Figure 4.** An example of the application of a neural network for atom finding (NN-1) and defect finding (NN-2).

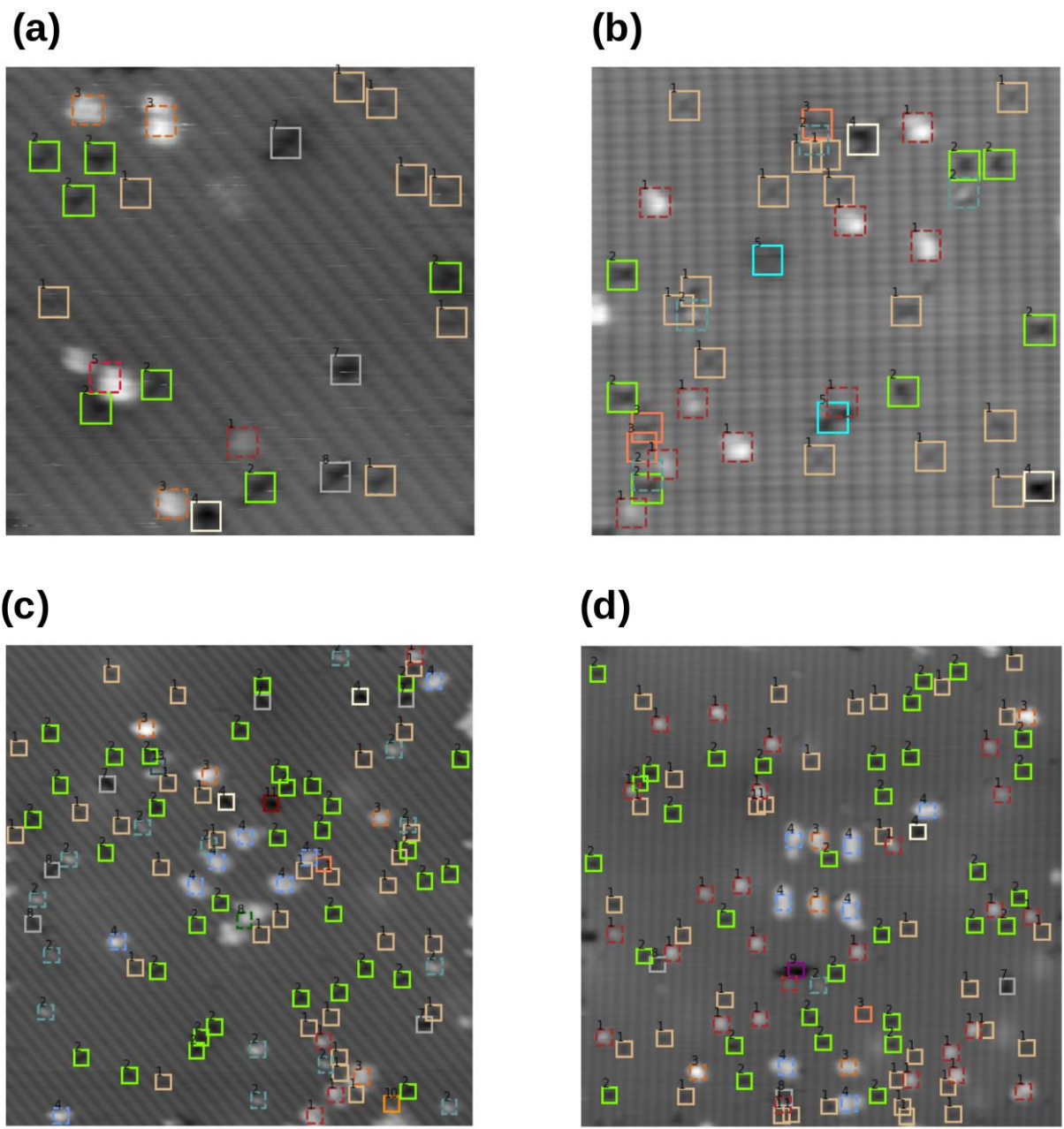

**FIGURE 5.** Decoded experimental images. Different color boxes annotated with different numbers indicate different classes (note that we have separate enumeration for "depressions" and "protrusions" and that we use solid and dotted lines for denoting "depressions" and "protrusions", respectively). For the "depressions", the first two classes (light brown (1) and green (2) boxes) comprise 81% of all the categorized depressions. For the protrusions, the class associated with lithographic (blue (4) and chocolate (1) boxes) and spurious (brown (3) boxes)

depassivation constitute 27% and 47% of all the categorized protrusions, whereas 20% (denoted by dark cyan(2) boxes) may be associated with water structures (see also Figure 6).

The outputs of the NN-1 and NN-2 networks for atom finding and defect finding are combined and used for the automated engineering of defect features which are then fed into the mean-shift clustering (msc) algorithm for the unsupervised classification of surface defects (see schematics in Figure 2). Notice that this approach is somewhat non-standard for classical machine learning applications and is heavily rooted in domain-specific physics. Specifically, this weakly supervised approach combines the features of classical unsupervised (in terms of feature learning) deep learning with "standard" machine learning which require feature engineering. The outputs of the two networks are then combined and used for the automated engineering of defect features (Figure 2). Specifically, we use the following features i) the average values of apparent height in the centers of the detected defects normalized by the average height of the detected atomic centers, ii) the area of the blobs associated with defects in the NN-2 output normalized by the average unit cell area, iii) the circularity of blobs associated with defects in the NN-2 output. We note that more features can be added in principle. The constructed features are then used to train a mean shift clustering algorithm[33,34], which does not require to manually input the number of clusters, for unsupervised classification of the surface defects (Figure 5 and Figure 6). This approach allows classifying atomic defects on Si surfaces and creating libraries of surface atomic defects, which can be later refined by domain experts and used to train various different machine learning classifiers (see schematics in Fig. 2).

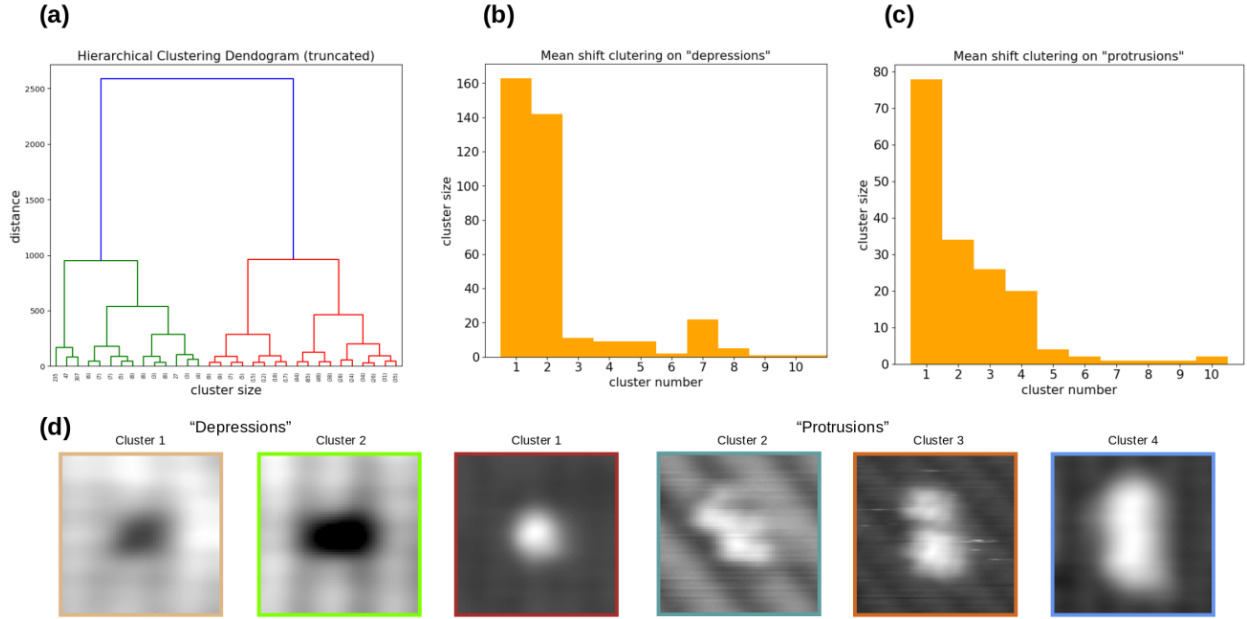

**FIGURE 6.** Clustering analysis of defect features. (a) Hierarchical clustering dendrogram analysis of all the features. (b, c) Number of defects in each class after performing mean-shift clustering analysis separately for "depressions" (b) and "protrusions" (c). (d) Averaged images of the corresponding clusters. Clusters 1 and 2 in the "depression" category and clusters 1 and 4 in the "protrusion" category are averaged from the defects identified for a single image. For cluster 4, only those defects were used whose major axis aligned with the direction of the slow scan. Cluster 2 and 3 in "the protrusion" category were identified from a separate image.

The results of the unsupervised classification with a mean-shift clustering algorithm trained on ~550 defects are shown for several experimental images in Figure 5. Our approach also allows a human operator to correct/refine the results of unsupervised classification, as well as (optionally) using them as an input to a supervised classification scheme. Because the mean-shift clustering was performed on the engineered features and the images in the dataset have different size and sample/scan rotation, it is not possible to visualize each cluster "centroids" (that is, pull an averaged image representing each defect class). However, from inspecting individual images (Figure 5) and the total number of defects in each class (Figure 6b, c), it is very easy to see that most of the depressions falls into one of two categories: one is of structural origin associated with missing lattice atoms (green boxes) and the other likely due to a formation of 'dihydride' structure and the associated changes in the local density of states (light brown boxes) (see also Fig. 6d). For protrusions, most of the defects falls into four major categories. Specifically, the defect classes associated with structures produced by lithographic depassivation are seen as blue and chocolate boxes, while the brown boxes are interpreted as "spurious depassivation". The dark cyan boxes denote structures potentially associated with water (see also Fig. 6d).

In terms of fabricating atomic-scale devices, identification of the location of these various classes is very important. The three-dimer patterns mentioned above in Figure 1 would need to avoid both types of depression features, while bigger patterns such as source or drain electrodes would not be affected. However, while protrusions identified as adatoms would also need to be avoided, those identified as missing H atoms or spurious depassivation can be written over without consequence. Automatically identifying suitable and unsuitable areas for writing a device pattern, a step which would normally require a human operator, is important progress towards full automation of atomic device fabrication. Larger features such as step edges are important to identify, as the dimer row directions rotate 90° on adjacent terraces, which would have consequences for the direction of tip vectors for the pattern writing. Moreover, there is a ¼ pixel shift in the local position of the dimer rows from terrace to terrace which would need to be accounted for in the fine control of tip position for the most precise writing.

We further note that atomically precise patterning is only possible when tip vectors during lithography are well aligned to the dimer rows on the Si (100) 2x1 surface. The identification of step edges and therefor atomic terraces is critical to assuring that excellent alignment to the Si (100) 2x1 lattice is possible. Similarly, without terrace and step edge identification alignment to surface lattice is impossible. The use of Fourier Transform analysis for alignment to the lattice requires a reasonably large area scan to assure accurate identification of the dimer row angle and position. Hence, accurate detection of atoms in dimer rows could be used in place of the Fourier Transform analysis for quicker more accurate identification of the surface lattice.

Overall, automated identification of surface defects will be extremely valuable for assessing surface quality after sample preparation, determining the suitability of a given area for a specific pattern, optimizing exposure conditions for lithography, alignment to fiducial marks or previously written features for accurate pattern placement, metrology to determine lithography accuracy, automated inspection and correction of the lithography process, and eventually for controlling multiple scanners in a highly parallel array. Especially the latter stage is impossible without automated image analytics. Past the simple image-based identification processes, we see machine learning as an automated process optimization tool for faster more accurate atomic scale patterning in the face of inevitable changes in tip and surface conditions.

In conclusion, we have developed a multiscale multi feature deep learning-based workflow for automatic segmentation and defect identification in the STM images. This workflow is tailored to fragment the images into the atomically-flat terrace regions and in parallel detect the point defects and atomic features. This "parallelization" yields a robust method that can be used for development of automated experimentation and atomic manipulation in STM. Here, we further demonstrate the automated creation of a defect library on Si(100) surface, from which information on the surface preparation and chemistry can be inferred. In the future, we are planning to develop theory assisted machine learning algorithms that allow internal image correction using the ideal Si(100) images to recover the tip state and factor in this information in the classifier.

**Methods:**

All the deep learning networks were implemented in PyTorch deep learning framework. The NN-0 and NN-1 were based on a modified U-Net model architecture[32] where regular convolutions in the middle ("bottleneck") block were replaced with a spatial pyramid of dilated convolutions with dilation rates {2, 4, 6}. The NN-2 consisted of three convolutional layers, each with 25 convolutional filters, followed by max-pooling operation, two back-to-back blocks with spatial pyramids of dilated convolutions with rates {2, 4, 6} and 50 filters in each layer, the bilinear upsampling operation and the three more regular convolutions with the same number of filters as in the first three layers. The size of all the convolutional filters was 3-by-3 and all activations were leaky ReLUs with a negative slope of 0.01 in all the three networks. The connected components labeling (for NN-0) and the center of the mass assignment (for NN-1 and NN-2) were performed using multidimensional image processing package scipy.ndimage. The analysis of atomic and defect contours for feature engineering was performed using OpenCV package.

**Acknowledgements**

This work was supported by Department of Energy's Small Business Innovation Research Grant (UF, JO, JR, MZ). Part of the work was performed at the Oak Ridge National Laboratory's Center for Nanophase Materials Sciences (CNMS), a U.S. Department of Energy, Office of Science User Facility (SVK).

This manuscript has been authored by UT-Battelle, LLC, under Contract No. DE-AC05-00OR22725 with the U.S. Department of Energy. The United States Government retains and the publisher, by accepting the article for publication, acknowledges that the United States Government retains a non-exclusive, paid-up, irrevocable, world-wide license to publish or reproduce the published form of this manuscript, or allow others to do so, for United States Government purposes. The Department of Energy will provide public access to these results of federally sponsored research in accordance with the DOE Public Access Plan (http://energy.gov/downloads/doe-public-access-plan).
14

**References**:

1. Richard Feynman talk: "There is Plenty of Room at the Bottom"
https://www.zyvex.com/nanotech/feynman.html

2. Drexler, K. E. (1986). *Engines of creation*. Garden City, N.Y: Anchor Press/Doubleday

3. Eigler, D. M. & Schweizer, E. K. Positioning single atoms with a scanning tunnelling microscope. *Nature* **344**, 524–526 (1990).

4. The 15th Anniversary of the U.S. National Nanotechnology Initiative. *ACS Nano* **12**, 10567–10569 (2018).

5. Crommie, M. F., Lutz, C. P. & Eigler, D. M. Confinement of electrons to quantum corrals on a metal surface. *Science* **262**, 218–220 (1993).

6. Moon, C. R., Mattos, L. S., Foster, B. K., Zeltzer, G. & Manoharan, H. C. Quantum holographic encoding in a two-dimensional electron gas. *Nat. Nanotechnol.* **4**, 167–172 (2009).

7. Heinrich, A. J., Lutz, C. P., Gupta, J. A. & Eigler, D. M. Molecule cascades. *Science* **298**, 1381–1387 (2002).

8. Loth, S., Baumann, S., Lutz, C. P., Eigler, D. M. & Heinrich, A. J. Bistability in atomic-scale antiferromagnets. *Science* **335**, 196–199 (2012).

9. Natterer, F. D. *et al.* Reading and writing single-atom magnets. *Nature* **543**, 226–228 (2017).

10. Kiraly, B. *et al.* An orbitally derived single-atom magnetic memory. *Nat. Commun.* **9**, 3904 (2018).

11. Gerhard, L. *et al.* Magnetoelectric coupling at metal surfaces. *Nat. Nanotechnol.* **5**, 792–797 (2010).

12. Paul, W. *et al.* Control of the millisecond spin lifetime of an electrically probed atom. *Nat. Phys.* **13**, 403 (2016).

13. Boy, A. & Atom, H. The World's Smallest Movie. *IBM Company. [(accessed on 10 February 2018)]*

14. Devoret, M. H. & Schoelkopf, R. J. Superconducting circuits for quantum information: an outlook. *Science* **339**, 1169–1174 (2013).

15. Bruzewicz, C. D., Chiaverini, J., McConnell, R. & Sage, J. M. Trapped-ion quantum computing: Progress and challenges. *Applied Physics Reviews* **6**, 021314 (2019).

16. Pakkiam, P. *et al.* Single-Shot Single-Gate rf Spin Readout in Silicon. *Phys. Rev. X* **8**, 041032 (2018).

17. Pla, J. J. *et al.* A single-atom electron spin qubit in silicon. *Nature* **489**, 541–545 (2012).

18. Lu, T.-M. *et al. Engineering Spin-Orbit Interaction in Silicon*. (Sandia National Lab.(SNL-NM), Albuquerque, NM (United States), 2018).

19. Ramanayaka, A. N. *et al.* STM patterned nanowire measurements using photolithographically defined implants in Si(100). *Sci. Rep.* **8**, 1790 (2018).

20. Dyck, O., Kim, S., Kalinin, S. V. & Jesse, S. Placing single atoms in graphene with a scanning transmission electron microscope. *Appl. Phys. Lett.* **111**, 113104 (2017).

21. Susi, T., Meyer, J. C. & Kotakoski, J. Manipulating low-dimensional materials down to the level of single atoms with electron irradiation. *Ultramicroscopy* **180**, 163–172 (2017).

22. Dyck, O. *et al.* Atom-by-atom fabrication with electron beams. *Nature Reviews Materials* 1 (2019).

23. Jesse, S. *et al.* Direct atomic fabrication and dopant positioning in Si using electron beams with active real-time image-based feedback. *Nanotechnology* **29**, 255303 (2018).

24. Ziatdinov, M. *et al.* Deep Learning of Atomically Resolved Scanning Transmission Electron Microscopy Images: Chemical Identification and Tracking Local Transformations. *ACS Nano* **11**, 12742–12752 (2017).

25. Madsen, J. *et al.* A Deep Learning Approach to Identify Local Structures in Atomic-Resolution Transmission Electron Microscopy Images. *Advanced Theory and Simulations* **1**, (2018).

26. Maksov, A., Dyck, O., Wang, K. & Xiao, K. Deep learning analysis of defect and phase evolution during electron beam-induced transformations in $WS_2$. *npj Computational Materials* (2019).

27. Ziatdinov, M. *et al.* Building and exploring libraries of atomic defects in graphene: scanning transmission electron and scanning tunneling microscopy study. *Science Advances* 5, eaaw8989 (2019).

28. Ziatdinov, M., Dyck, O., Jesse, S. & Kalinin, S. V. Atomic mechanisms for the Si atom dynamics in graphene: chemical transformations at the edge and in the bulk. *Advanced Functional Materials* 29 (52), 1904480 (2019).

29. Rashidi, M. & Wolkow, R. A. Autonomous Scanning Probe Microscopy in Situ Tip

Conditioning through Machine Learning. *ACS Nano* (2018). doi:10.1021/acsnano.8b02208

30. Rashidi, M. *et al.* Autonomous Atomic Scale Manufacturing Through Machine Learning. *arXiv [cond-mat.mtrl-sci]* (2019).

31. J.N. Randall, J.B. Ballard, J.W. Lyding, S. Schmucker, J.R. Von Ehr, R. Saini, H. Xu, and Y. Ding, Microelectron. Eng. **87**, 955 (2010).

32. Ronneberger, O., Fischer, P. & Brox, T. U-Net: Convolutional Networks for Biomedical Image Segmentation. *arXiv [cs.CV]* (2015).

33. Yizong Cheng. Mean shift, mode seeking, and clustering. *IEEE Trans. Pattern Anal. Mach. Intell.* **17**, 790–799 (1995).

34. Comaniciu, D. & Meer, P. Mean shift: A robust approach toward feature space analysis. *IEEE Trans. Pattern Anal. Mach. Intell.* 603–619 (2002).